\def\rfr#1{eq. (\ref{#1})}

\def\virg#1{``#1''}

\def\dert#1#2{\frac{{{d}}{#1}}{{{d}}{#2}}}              % derivate parziali e totali prima e seconda

\def\dert#1#2{\frac{{{d}}{#1}}{{{d}}{#2}}}              % derivate parziali e totali prima e seconda

\def\bar{\begin{eqnarray}}
\def\ear{\end{eqnarray}}
\def\bb{\bibitem}
\def\eqi{\begin{equation}}
\def\eqf{\end{equation}}
\def\eqia{\begin{eqnarray}}
\def\eqfa{\end{eqnarray}}
\def\rp#1#2{\frac{#1}{#2}}
\def\ct#1{\cite{#1}}
\def\lb#1{\label{#1}}

\documentclass[
numberedheadings % uncomment this option for numbered sections
  ]
  {aipproc}

\layoutstyle{8x11single}

\begin{document}

\title{The Impact of the External Field Effect in the MOdified Newtonian Dynamics on Solar System's Orbits}

\classification{\texttt{04.50.Kd; 04.80.Cc; 95.10.Ce; 95.10.Km; 96.50.Hp}}
\keywords      {Modified theories of gravity; Experimental tests of gravitational theories; Celestial mechanics; Ephemerides, almanacs, and calendars; Oort cloud
 }

\author{L. Iorio}{
  address={Viale Unit\`{a} di Italia 68, 70125, Bari (BA), Italy. INFN-Sezione di Pisa}
}

\begin{abstract}
 We looked at the orbital motions of test particles according to the External Field Effect (EFE) predicted by  the MOdified Newtonian Dynamics (MOND) in  the Oort cloud which falls in the deep MONDian regime ($r\approx 50-150$ kAU).
Concerning the interpolating function $\mu(x)$, we extensively used the forms $\mu_1=1/(1+x),\mu_2=x/(1+x^2)^{1/2},\mu_{3/2}=x/(1+x^{3/2})^{2/3}$.
We integrated both the MOND and the Newtonian equations of motion in Cartesian coordinates sharing the same initial conditions. We considered both ecliptic and nearly polar trajectories, all with high eccentricities ($e>0.1$). In order to evaluate the characteristic MOND parameters $\mu_g$ and $L_g$ entering the problem, we used two different values ($V=220$ km s$^{-1}$ and $V=254$ km s$^{-1}$) of the circular speed of the solar system's motion through the Galaxy; $V$ allows to evaluate  the Milky Way's gravitational field at the Sun's location. It turns out that EFE induces strong distortions of the Newtonian ellipses, especially in the ecliptic plane yielding more involved paths which span less extended spatial regions.
\end{abstract}

\maketitle

%%%%%%%%%%%%%%%%%%%%%%%%%%%%%%%%%%%%%%%%%%%%
%% MAINMATTER
%%%%%%%%%%%%%%%%%%%%%%%%%%%%%%%%%%%%%%%%%%%%

\section{Introduction}
MOdified Newtonian Dynamics (MOND) \ct{Mil83a}  was proposed by Milgrom in 1983 to explain  the  discrepancy between the observed kinematics of the exterior parts of spiral galaxies and the predicted one on the basis of the Newtonian dynamics and the baryonic matter detected from the emitted electromagnetic radiation (visible stars and gas clouds) \ct{Bos,Rub} without invoking exotic forms of still undetected non-baryonic Dark Matter. MOND postulates that
for systems  experiencing total gravitational acceleration  $A < A_0$, with \ct{Bege} \eqi A_0= (1.2\pm 0.27)\times 10^{-10} \ {\rm m\ s}^{-2},\eqf
\eqi \vec{A}\rightarrow \vec{A}_{\rm MOND}=-\rp{\sqrt{A_0GM}}{r}{\hat{r}}.\lb{MOND}\eqf   More precisely, it holds \eqi A = \rp{A_{\rm Newton}}{\mu(x)},\ x\equiv\rp{A}{A_0};\lb{appromond1}\eqf
$\mu(x)\rightarrow 1$ for $x\gg 1$, i.e. for large accelerations (weak MOND regime, strong Newtonian regime), while $\mu(x)\rightarrow x$ yielding \rfr{MOND} for $x\ll 1$, i.e. for small accelerations (deep MOND regime).
The most widely used forms for the interpolating function $\mu$ are \cite{Fam,joint}
\begin{eqnarray}
% \nonumber to remove numbering (before each equation)
  \mu_1(x) &=& \rp{x}{1+x},\lb{mu1}\\
  \mu_2(x) &=& \rp{x}{(1+x^2)^{1/2}}.\lb{mu2}
\end{eqnarray}
Such forms, and also another one, as we will see later, can be reduced to the following high-acceleration limit ($x\gg 1$)
\eqi \mu\approx 1-k_0 x^{-\alpha}.\eqf Indeed, \rfr{mu1} corresponds to $k_0=1$, $\alpha=1$, while \rfr{mu2} corresponds to $k_0=1/2$ and $\alpha=2$.
%It recently turned out that the simpler form of \rfr{mu1} yields  better results in fitting the terminal velocity curve of the Milky Way, the rotation curve of %the standard external galaxy NGC 3198 \cite{Fam,kazzo,scassa} and of a sample of 17 high
%surface brightness, early-type disc galaxies \cite{Noor}.

In the framework of MOND, the internal dynamics of a
gravitating system s embedded in a larger one S is affected
by the external background field $\vec{E}$ of S, even if
it is constant and uniform, thus implying a violation
of the Strong Equivalence Principle: it is the so-called
External Field Effect (EFE). In the case of the solar
system, $E$ would be $A_{\rm cen}\approx 10^{-10}$ m s$^{-2}$ because of its
motion through the Milky Way \cite{Mil83a,joint,Milpazzo}.
\section{EFE in the planetary regions of the solar system}
Perhaps the absence of explicit expressions for the action of EFE in the weak MOND regime, i.e. in the planetary regions of the solar system, may have lead to a misunderstanding in some sectors of the scientific community. Indeed, a researcher active in MOND writes: \virg{MOND breaks down the Strong Equivalence Principle. This means that the acceleration of solar system's bodies depends indeed on the background gravitational field and not only on the tidal field. As shown by Milgrom, even if the
external field was constant (and the tidal force vanishes), the internal acceleration would depend on the external field. Claiming that $A_{\rm cen}$ is irrelevant is only valid if the field equation were linear.} He/she also adds that \virg{for trans-Neptunian objects and planets, one can ignore the $A_{\rm cen}$.}
Another researcher working on MOND tries to go in deeper details  by  writing: \virg{For the main planets, the acceleration
is much larger than $A_0$ (the order of magnitude of the EFE), and the effect is negligible [...]
The EFE maintains a constant direction in the planet revolution, and its effect cancels out. }
Such statements are likely expressions of a widely diffuse belief about EFE in solar system.

In fact, it has been shown in \cite{Ior} that a constant and uniform acceleration $\vec{E}=E_x \hat{x} + E_y \hat{y} +E_z \hat{z}$ directed along a fixed direction $\hat{n}$ and with $E=A_{\rm cen}\approx A_0$ does induce non-zero long-period, i.e. averaged over one orbital revolution, effects on the Keplerian orbital elements of a planet. Moreover, the resulting perihelion precessions of the inner planets would be $4-6$ orders of magnitude larger than the present-day limits on the recently estimated non-standard perihelion rates \cite{Pit05}.

Milgrom \cite{Mil09}  recently made a step forward by explicitly working out EFE in the deep Newtonian regime. It turns out to be equivalent
to the action of  a distant, localized body X \cite{IorX}
\eqi \vec{A}_{\rm X} \approx -\mathcal{K}\vec{r} + 3\mathcal{K}\left(\vec{r}\cdot\hat{n}_{\rm X}\right)\hat{n}_{\rm X},\lb{acce}\eqf
where
\eqi\mathcal{K}\equiv \rp{GM_{\rm X}}{r^3_{\rm X}}\eqf
is the so-called tidal parameter of X and $\hat{n}_{\rm X}$ is a constant unit vector in the direction of X.
The acceleration of \rfr{acce} comes from the following quadrupolar potential \cite{Hog91}
\eqi U_{\rm X}\approx\rp{\mathcal{K}}{2}[r^2 -3(\vec{r}\cdot\hat{n}_{\rm X})^2].\lb{pot}\eqf
The  result by Milgrom can be obtained with the replacement
\eqi \mathcal{K}\rightarrow -\rp{q}{2}\left(\rp{A_0}{r_t}\right),\ r_t=\sqrt{\rp{GM_{\odot}}{A_0}}=6.833\ {\rm kAU},\eqf
where $-q$ is the MOND quadrupolar parameter.
Latest data from Saturnian perihelion \cite{Pit,Fie} yield \cite{IorX}
\eqi-q\geq 0.2\eqf with a maximum value
\eqi -q_{\rm max}=0.4-1.\eqf
However, the hypothesis that a planetary-sized body X really exists in the distant regions of the solar system should be regarded as a serious competitor because its confrontation with the planetary perihelion rates \cite{IorX}  yields plausible
values for the distance of X for various choices of its mass, in agreement with previous theoretical predictions for X made to explain certain features of the Kuiper Belt and the Oort cloud.
\section{Orbits of Oort comets in MOND}
Moving to the deep MOND regime in the remote periphery of the solar system, let us define the following quantities
\begin{eqnarray}
% \nonumber to remove numbering (before each equation)
  \eta &=& \rp{A_{\rm cen}}{A_0}\geq 1, \\
  L &=& \rp{x}{\mu}\left(\dert \mu x\right), \\
  \mu_g &=& \mu(\eta), \\
  L_g &=& L(\eta).
\end{eqnarray}
In the weak acceleration regime, for
\eqi r\gg {r_t}\eta^{-1/2},\eqf i.e. in the Oort cloud \cite{Oo}, the  action of EFE is different, so that the total acceleration felt by an Oort comet is \cite{joint,Mil09}
\eqi \vec{A} = -\rp{GM}{\mu_g(1+L_g)^{1/2}}\left(\rp{x^2}{1+L_g} + y^2 + z^2\right)^{-3/2}\left(
                                                                                           \begin{array}{c}
                                                                                             \rp{x}{1+L_g} \\
                                                                                             y \\
                                                                                             z \\
                                                                                           \end{array}
                                                                                         \right).
\eqf
Note that, since the ecliptic longitude and latitude of the Galactic Center are about $\lambda_{\rm GC}\approx 180$ deg, $\beta_{\rm GC}\approx -6$ deg, EFE is directed along the $X$ axis of the ICRF, i.e. the barycentric frame in which the motion of solar system's objects are usually studied.

Concerning $L$, we have
\begin{eqnarray}
% \nonumber to remove numbering (before each equation)
  \mu_1 = \rp{x}{1+x} &\rightarrow & L_1 =\rp{1}{1+x}, \\
  \mu_2 = \rp{x}{(1+x^2)^{1/2}} &\rightarrow & L_2 = \rp{1}{1+x^2},\\
  \mu_{3/2} = \rp{x}{(1+x^{3/2})^{2/3}}&\rightarrow & L_{3/2} = \rp{1}{1+x^{3/2}}.
\end{eqnarray}
The form $\mu_{3/2}$ was proposed in \cite{Mil09}.
The value of $\eta$ depends on the Galactic field at the solar system's location which can be obtained from its centrifugal acceleration
\eqi A_{\rm cen}=\rp{V^2}{R},\eqf where $V$ is the speed of the Local Standard of Rest (LSR) and $R=8.5$ kpc is the Galactocentric distance. The standard IAU value for  the speed is $V=220$ km s$^{-1}$, but recent determinations \cite{Reid} obtained with the Very Long Baseline Array and the Japanese VLBI Exploration of Radio Astronomy project yield a higher value: $V=254\pm 16$ km s$^{-1}$.
Thus, $\eta$ ranges from 1.5 to 2.3.
\subsection{Ecliptic orbits}
\subsubsection{Case $\mu_{3/2}$}
\begin{figure}[!h]
{ \centering \begin{tabular}{cc}
\includegraphics[height=0.3\textheight,width=0.5\columnwidth]{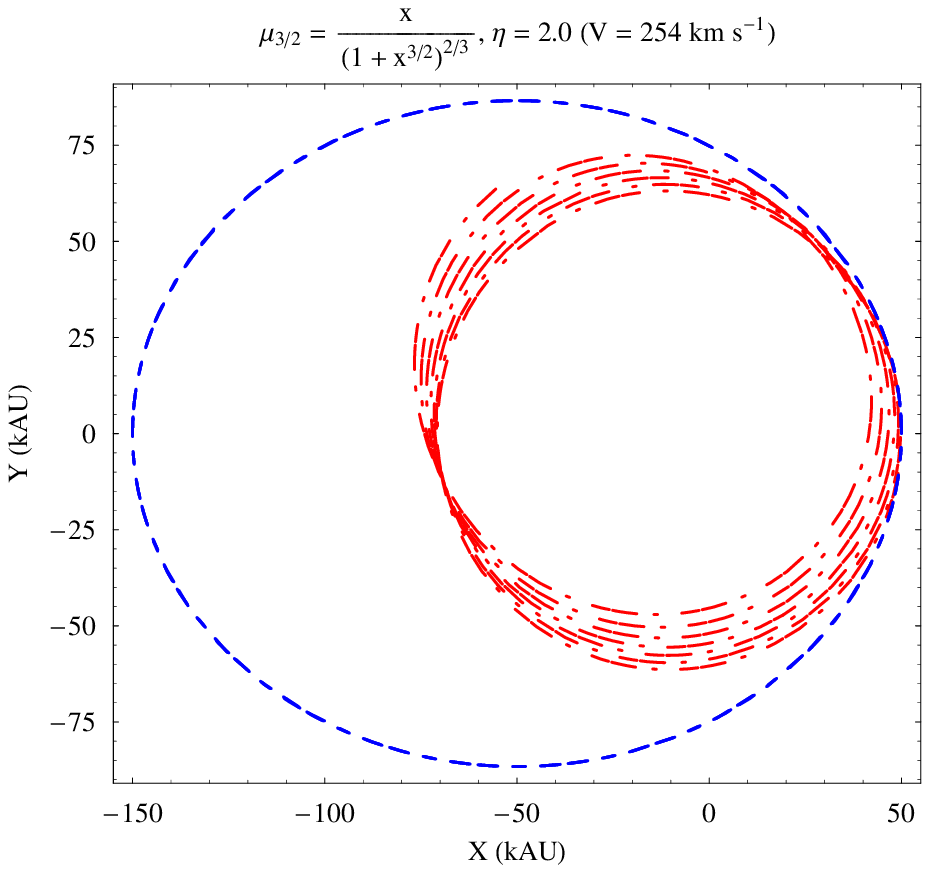}& %[height=0.3\textheight,width=0.4\columnwidth]{xy_polar_mu2_old.eps}&
\includegraphics[height=0.3\textheight,width=0.5\columnwidth]{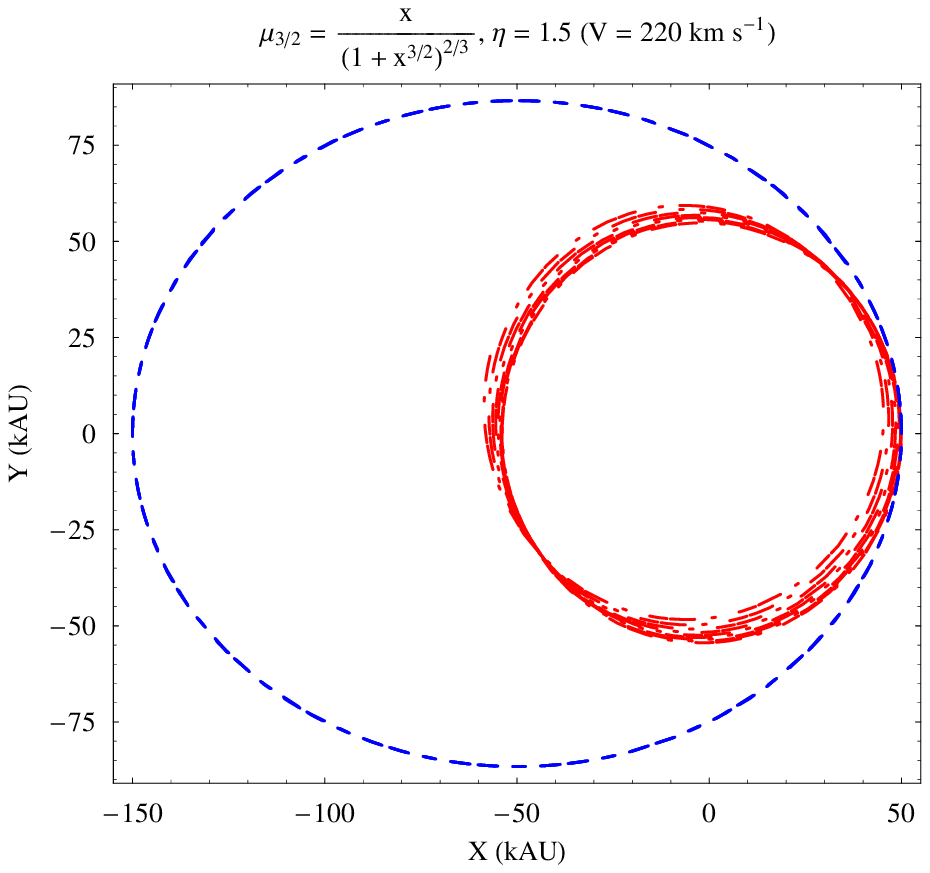}\\ %[height=0.3\textheight,width=0.3\columnwidth]{xz_polar_mu2_old.eps} &
\end{tabular} \par }
\caption{Numerically integrated orbits of an ecliptic Oort comet with $a=100$ kAU, $e=0.5$, $P_{\rm b}=31.6$ Myr. Dashed blue line: Newton. Dash-dotted red line: MOND with $\mu_{3/2}$. Left panel: $\eta=2.0$ corresponding to $V=254$ km s$^{-1}$. Right panel: $\eta=1.5$ corresponding to $V=220$ km s$^{-1}$. The initial conditions are $x_0=a(1-e), y_0=z_0=0,\dot x_0=0,\dot y_0=n a \sqrt{\rp{1+e}{1-e}},\dot z_0=0$. The time span of the integration is $-3P_{\rm b}\leq t\leq 0$.\label{uno}}
\end{figure}

We will, now, consider an Oort comet whose Newtonian orbit covers the entire extension of the Oort cloud. It has semimajor axis $a=100$ kAU and eccentricity $e=0.5$, so that its perihelion is at 50 kAU and its aphelion is at 150 kAU; for the sake of simplicity, we will assume it lies in the ecliptic plane. Its Newtonian orbital period is $P_{\rm b}=31.6$ Myr.
We will, first, use $\eta=2.0\ (V=254\ {\rm km\ s^{-1}})$ and $\mu_{3/2}$, so that
\begin{eqnarray}
% \nonumber to remove numbering (before each equation)
  \mu_g &=& 0.82 \\
  L_g &=& 0.25
\end{eqnarray}
Left panel of Figure \ref{uno} depicts the integrated Newtonian (dashed blue line) and MONDian (dash-dotted red line) orbits for the same initial conditions for $-3 P_{\rm b}\leq t\leq 0$.
%
%
%
%
%\begin{figure}
% \includegraphics[height=.25\textheight]{icsipsilon_3P_new}
%
%\caption{\footnotesize{Numerically integrated orbits of an Oort comet with $a=100$ kAU, $e=0.5$, $P_{\rm b}=31.6$ Myr. Dashed blue line: Newton. Dash-dotted %red line: MOND with $\mu_{3/2}$, $\eta=2.0$ ($V=254$ km s$^{-1}$). The initial conditions are $x_0=a(1-e), y_0=z_0=0,\dot x_0=0,\dot y_0=n a
%\sqrt{\rp{1+e}{1-e}},\dot z_0=0$. The time span of the integration is $-3 P_{\rm b}\leq t\leq 0$.}\label{uno}}
%\end{figure}
%
%
%
%
%
%
In the right panel of Figure \ref{uno} we show the case $\eta=1.5\ (V=220\ {\rm km\ s^{-1}})$ yielding
\begin{eqnarray}
% \nonumber to remove numbering (before each equation)
  \mu_g &=& 0.75 \\
  L_g &=& 0.34.
\end{eqnarray}
%
%
%\begin{figure}
% \includegraphics[height=.25\textheight]{icsipsilon_3P_old}
%
%\caption{\footnotesize{Numerically integrated orbits of an Oort comet with $a=100$ kAU, $e=0.5$, $P_{\rm b}=31.6$ Myr. Dashed blue line: Newton. Dash-dotted %red line: MOND with $\mu_{3/2}$, $\eta=1.5$ ($V=220$ km s$^{-1}$). The initial conditions are $x_0=a(1-e), y_0=z_0=0,\dot x_0=0,\dot y_0=n a %\sqrt{\rp{1+e}{1-e}},\dot z_0=0$. The time span of the integration is $-3 P_{\rm b}\leq t\leq 0$.}\label{due}}
%\end{figure}
%
%
The MOND trajectories are not closed and are much less spatially extended that the Newtonian ones; the overall shrinking of the orbit is more marked for the standard value of the LSR circular speed (right panel of Figure \ref{uno}).

\begin{figure}[!h]
{ \centering \begin{tabular}{c}
\includegraphics{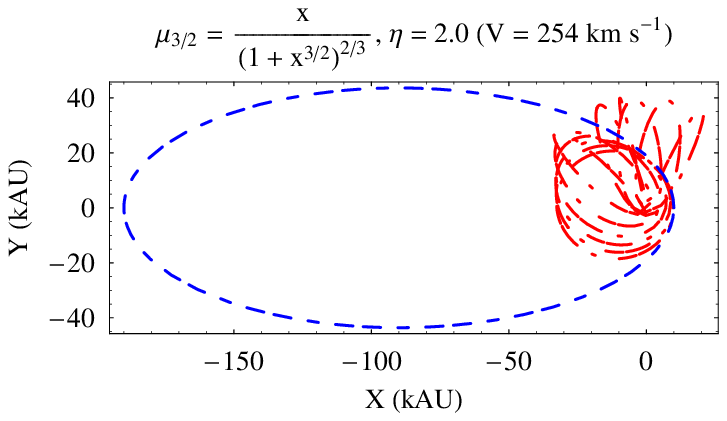}\\ %[height=0.3\textheight,width=0.4\columnwidth]{xy_polar_mu2_old.eps}&
\includegraphics{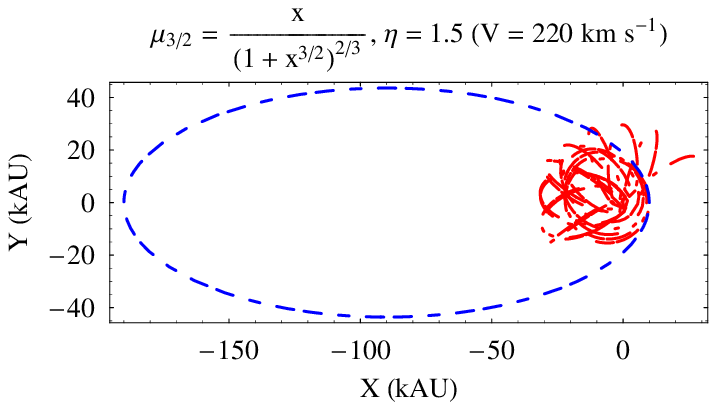}\\ %[height=0.3\textheight,width=0.3\columnwidth]{xz_polar_mu2_old.eps} &
\end{tabular} \par }
\caption{Numerically integrated orbits of an ecliptic Oort comet with $a=100$ kAU, $e=0.9$, $P_{\rm b}=31.6$ Myr. Dashed blue line: Newton. Dash-dotted red line: MOND with $\mu_{3/2}$. Upper panel: $\eta=2.0$ corresponding to $V=254$ km s$^{-1}$. Lower panel: $\eta=1.5$ corresponding to $V=220$ km s$^{-1}$. The initial conditions are $x_0=a(1-e), y_0=z_0=0,\dot x_0=0,\dot y_0=n a \sqrt{\rp{1+e}{1-e}},\dot z_0=0$. The time span of the integration is $-P_{\rm b}\leq t\leq 0$.\label{tre}}
\end{figure}

Such an effect is particularly notable for highly elliptical Newtonian orbits, as shown by Figure \ref{tre} for $e=0.9$ and $-P_{\rm b}\leq t\leq 0$.
%
%\begin{figure}
% \includegraphics[height=.25\textheight]{icsipsilon_3P_new_elli}
%
%\caption{\footnotesize{Numerically integrated orbits of an Oort comet with $a=100$ kAU, $e=0.9$, $P_{\rm b}=31.6$ Myr. Dashed blue line: Newton. Dash-dotted %red line: MOND with $\mu_{3/2}$, $\eta=2.0$ ($V=254$ km s$^{-1}$). The initial conditions are $x_0=a(1-e), y_0=z_0=0,\dot x_0=0,\dot y_0=n a %\sqrt{\rp{1+e}{1-e}},\dot z_0=0$. The time span of the integration is $-P_{\rm b}\leq t\leq 0$.}\label{tre}}
%\end{figure}
%
%
%
%\begin{figure}
% \includegraphics[height=.25\textheight]{icsipsilon_3P_old_elli}
%
%\caption{\footnotesize{Numerically integrated orbits of an Oort comet with $a=100$ kAU, $e=0.9$, $P_{\rm b}=31.6$ Myr. Dashed blue line: Newton. Dash-dotted %red line: MOND with $\mu_{3/2}$, $\eta=1.5$ ($V=220$ km s$^{-1}$). The initial conditions are $x_0=a(1-e), y_0=z_0=0,\dot x_0=0,\dot y_0=n a %\sqrt{\rp{1+e}{1-e}},\dot z_0=0$. The time span of the integration is $-P_{\rm b}\leq t\leq 0$.}\label{quattro}}
%\end{figure}
%
Note also how wildly the MOND trajectory changes during one Keplerian orbital period.
\subsubsection{Case $\mu_2$}
\begin{figure}[!h]
{ \centering \begin{tabular}{cc}
\includegraphics[height=0.3\textheight,width=0.5\columnwidth]{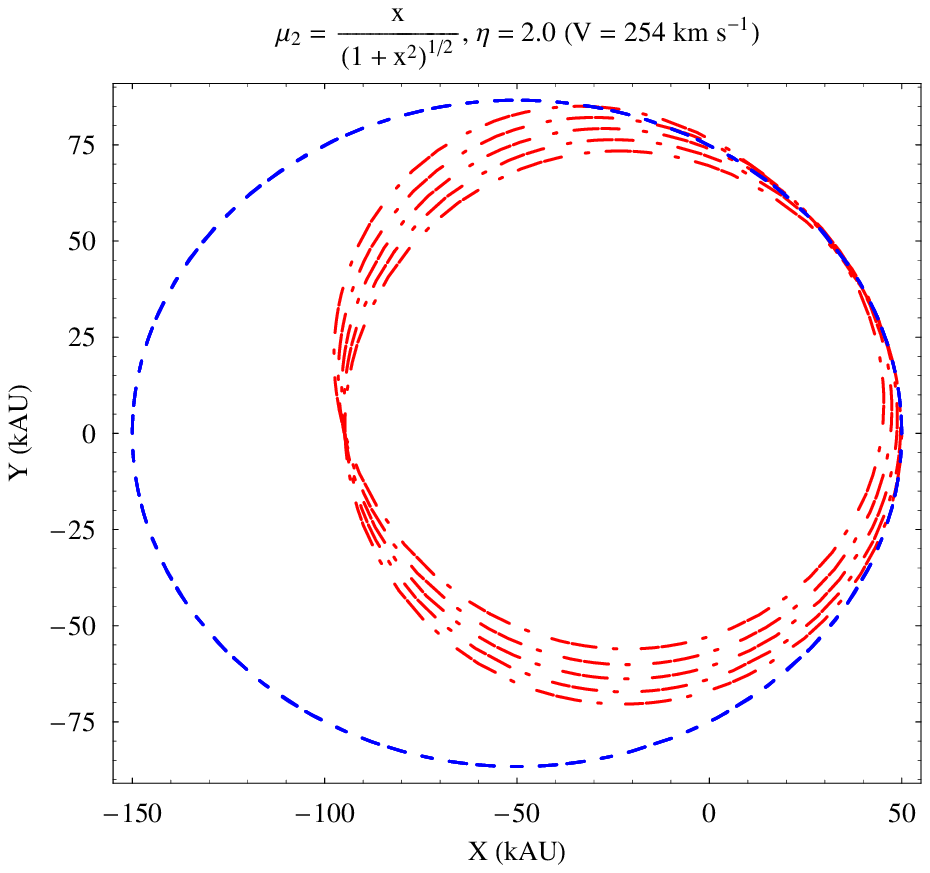}& %[height=0.3\textheight,width=0.4\columnwidth]{xy_polar_mu2_old.eps}&
\includegraphics[height=0.3\textheight,width=0.5\columnwidth]{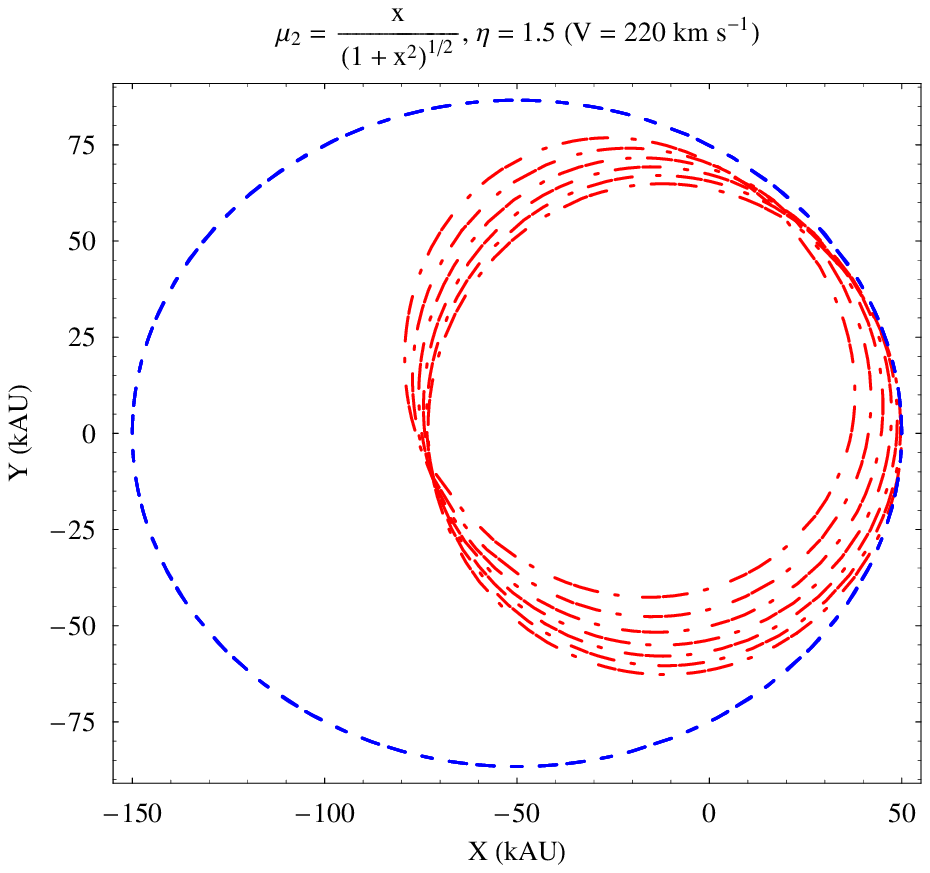}\\ %[height=0.3\textheight,width=0.3\columnwidth]{xz_polar_mu2_old.eps} &
\end{tabular} \par }
\caption{Numerically integrated orbits of an ecliptic Oort comet with $a=100$ kAU, $e=0.5$, $P_{\rm b}=31.6$ Myr. Dashed blue line: Newton. Dash-dotted red line: MOND with $\mu_{2}$. Left panel: $\eta=2.0$ corresponding to $V=254$ km s$^{-1}$. Right panel: $\eta=1.5$ corresponding to $V=220$ km s$^{-1}$. The initial conditions are $x_0=a(1-e), y_0=z_0=0,\dot x_0=0,\dot y_0=n a \sqrt{\rp{1+e}{1-e}},\dot z_0=0$. The time span of the integration is $-3P_{\rm b}\leq t\leq 0$.\label{cinque}}
\end{figure}

\begin{figure}[!h]
{ \centering \begin{tabular}{c}
\includegraphics{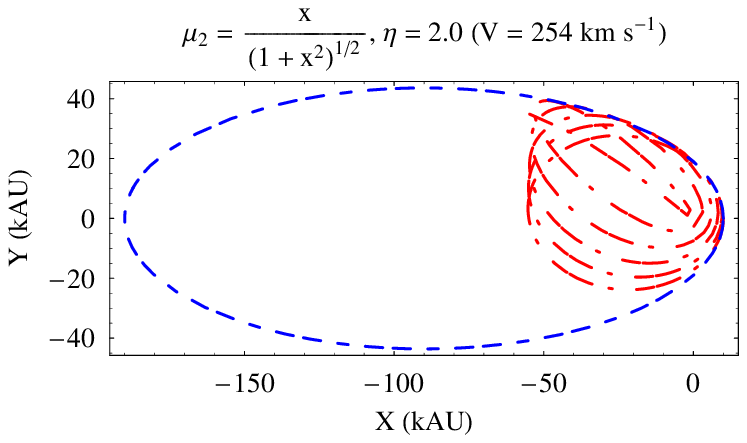}\\ %[height=0.3\textheight,width=0.4\columnwidth]{xy_polar_mu2_old.eps}&
\includegraphics{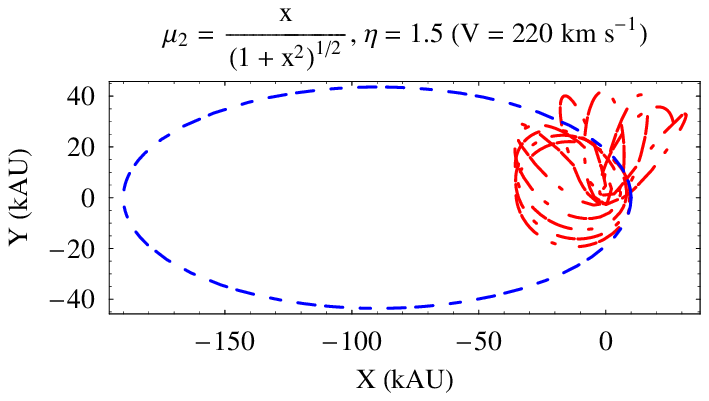}\\ %[height=0.3\textheight,width=0.3\columnwidth]{xz_polar_mu2_old.eps} &
\end{tabular} \par }
\caption{Numerically integrated orbits of an ecliptic Oort comet with $a=100$ kAU, $e=0.9$, $P_{\rm b}=31.6$ Myr. Dashed blue line: Newton. Dash-dotted red line: MOND with $\mu_{2}$. Upper panel: $\eta=2.0$ corresponding to $V=254$ km s$^{-1}$. Lower panel: $\eta=1.5$ corresponding to $V=220$ km s$^{-1}$. The initial conditions are $x_0=a(1-e), y_0=z_0=0,\dot x_0=0,\dot y_0=n a \sqrt{\rp{1+e}{1-e}},\dot z_0=0$. The time span of the integration is $-P_{\rm b}\leq t\leq 0$.\label{sette}}
\end{figure}

In Figure \ref{cinque} ($e=0.5$) and Figure \ref{sette}  ($e=0.9$) we use $\mu_2$.
The values of the MONDian characteristic parameters are
\begin{eqnarray}
% \nonumber to remove numbering (before each equation)
  \mu_g &=& 0.89 \\
  L_g &=& 0.19
\end{eqnarray}
for $\eta=2.0$, and
\begin{eqnarray}
% \nonumber to remove numbering (before each equation)
  \mu_g &=& 0.83 \\
  L_g &=& 0.29
\end{eqnarray}
for $\eta=1.5$.
%
%\begin{figure}
% \includegraphics[height=.25\textheight]{icsipsilon_3P_sta_new}
%
%\caption{\footnotesize{Numerically integrated orbits of an Oort comet with $a=100$ kAU, $e=0.5$, $P_{\rm b}=31.6$ Myr. Dashed blue line: Newton. Dash-dotted %red line: MOND with $\mu_{2}$, $\eta=2.0$ ($V=254$ km s$^{-1}$). The initial conditions are $x_0=a(1-e), y_0=z_0=0,\dot x_0=0,\dot y_0=n a %\sqrt{\rp{1+e}{1-e}},\dot z_0=0$. The time span of the integration is $-3 P_{\rm b}\leq t\leq 0$.}\label{cinque}}
%\end{figure}
%
%
%
%
%\begin{figure}
% \includegraphics[height=.25\textheight]{icsipsilon_3P_sta_old}
%
%\caption{\footnotesize{Numerically integrated orbits of an Oort comet with $a=100$ kAU, $e=0.5$, $P_{\rm b}=31.6$ Myr. Dashed blue line: Newton. Dash-dotted %red line: MOND with $\mu_{2}$, $\eta=1.5$ ($V=220$ km s$^{-1}$). The initial conditions are $x_0=a(1-e), y_0=z_0=0,\dot x_0=0,\dot y_0=n a %\sqrt{\rp{1+e}{1-e}},\dot z_0=0$. The time span of the integration is $-3 P_{\rm b}\leq t\leq 0$.}\label{sei}}
%\end{figure}
%
%
%
%
%
%
%
%\begin{figure}
% \includegraphics[height=.25\textheight]{icsipsilon_3P_sta_new_elli}
%
%\caption{\footnotesize{Numerically integrated orbits of an Oort comet with $a=100$ kAU, $e=0.9$, $P_{\rm b}=31.6$ Myr. Dashed blue line: Newton. Dash-dotted red %line: MOND with $\mu_{2}$, $\eta=2.0$ ($V=254$ km s$^{-1}$). The initial conditions are $x_0=a(1-e), y_0=z_0=0,\dot x_0=0,\dot y_0=n a \sqrt{\rp{1+e}{1-e}},\dot %z_0=0$. The time span of the integration is $-P_{\rm b}\leq t\leq 0$.}\label{sette}}
%\end{figure}
%
%
%
%
%\begin{figure}
% \includegraphics[height=.25\textheight]{icsipsilon_3P_sta_old_elli}
%
%\caption{\footnotesize{Numerically integrated orbits of an Oort comet with $a=100$ kAU, $e=0.9$, $P_{\rm b}=31.6$ Myr. Dashed blue line: Newton. Dash-dotted red %line: MOND with $\mu_{2}$, $\eta=1.5$ ($V=220$ km s$^{-1}$). The initial conditions are $x_0=a(1-e), y_0=z_0=0,\dot x_0=0,\dot y_0=n a \sqrt{\rp{1+e}{1-e}},\dot %z_0=0$. The time span of the integration is $-P_{\rm b}\leq t\leq 0$.}\label{otto}}
%\end{figure}
%
%

Also with such a form of the interpolating function $\mu$, orbits that are highly eccentric in Newtonian dynamics are confined to  much smaller spatial regions in MOND and experience high-frequency variations over one Keplerian orbital period (Figure \ref{sette}). However, for $\mu_2$ the largest extension of the MOND trajectory occurs for $V=220$ km s$^{-1}$ ($\eta=1.5$).
\subsubsection{Case $\mu_1$}
\begin{figure}[!h]
{ \centering \begin{tabular}{cc}
\includegraphics[height=0.3\textheight,width=0.5\columnwidth]{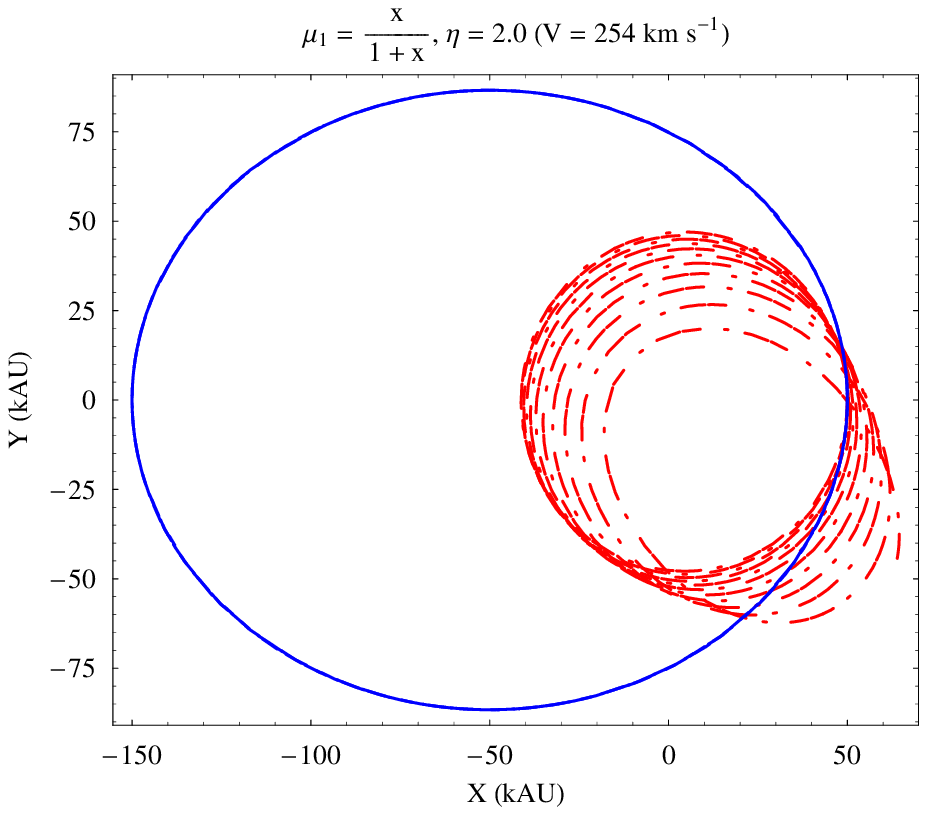}& %[height=0.3\textheight,width=0.4\columnwidth]{xy_polar_mu2_old.eps}&
\includegraphics[height=0.3\textheight,width=0.5\columnwidth]{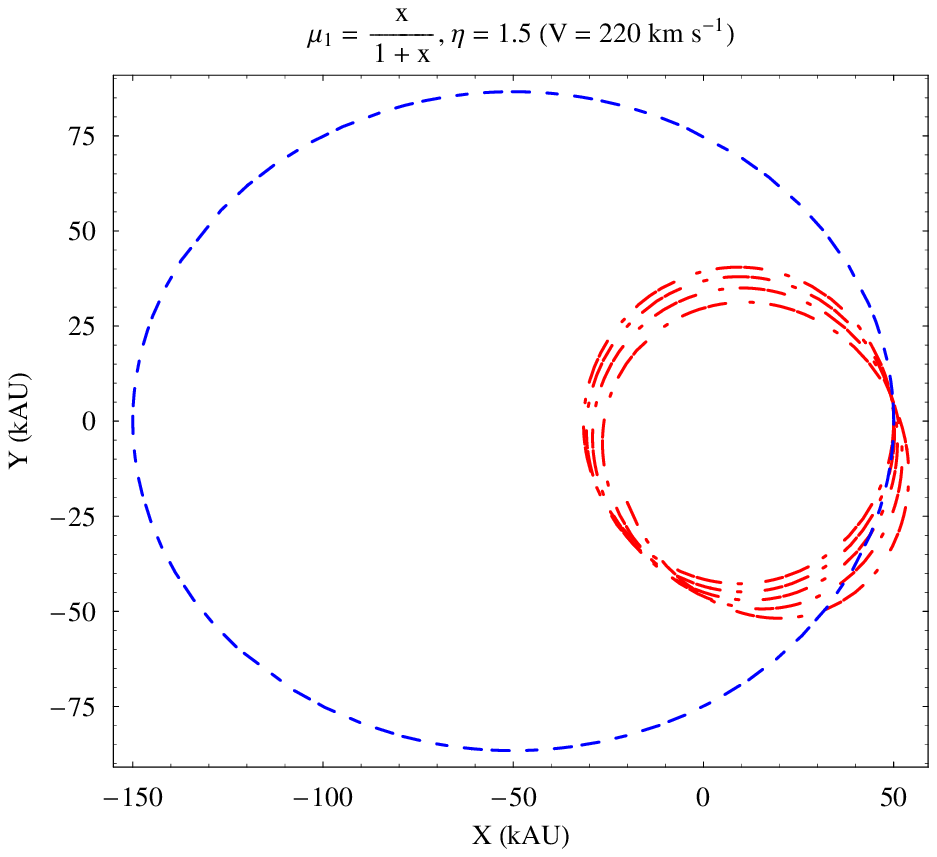}\\ %[height=0.3\textheight,width=0.3\columnwidth]{xz_polar_mu2_old.eps} &
\end{tabular} \par }
\caption{Numerically integrated orbits of an ecliptic Oort comet with $a=100$ kAU, $e=0.5$, $P_{\rm b}=31.6$ Myr. Dashed blue line: Newton. Dash-dotted red line: MOND with $\mu_{1}$. Left panel: $\eta=2.0$ corresponding to $V=254$ km s$^{-1}$. Right panel: $\eta=1.5$ corresponding to $V=220$ km s$^{-1}$. The initial conditions are $x_0=a(1-e), y_0=z_0=0,\dot x_0=0,\dot y_0=n a \sqrt{\rp{1+e}{1-e}},\dot z_0=0$. The time spans of the integration are $-3 P_{\rm b}\leq t\leq 0$ (left panel) and $-P_{\rm b}\leq t\leq 0$ (right panel).\label{nove}}
\end{figure}

Let us, now, examine the case $\mu_1$.
In the left panel of Figure \ref{nove} we show the  trajectory due to it of the Oort comet with $e=0.5$ over  $3 P_{\rm b}$ for $\eta=2.0$ which implies
\begin{eqnarray}
% \nonumber to remove numbering (before each equation)
  \mu_g &=& 0.67 \\
  L_g &=& 0.33.
\end{eqnarray}

\begin{figure}[!h]
{ \centering \begin{tabular}{c}
\includegraphics{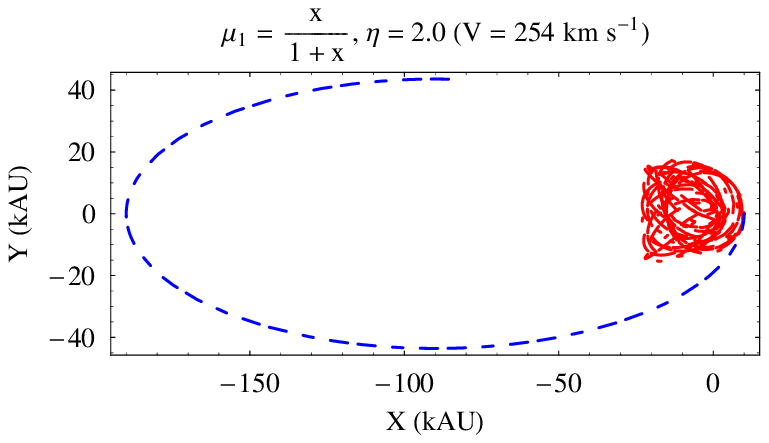}\\ %[height=0.3\textheight,width=0.4\columnwidth]{xy_polar_mu2_old.eps}&
\includegraphics{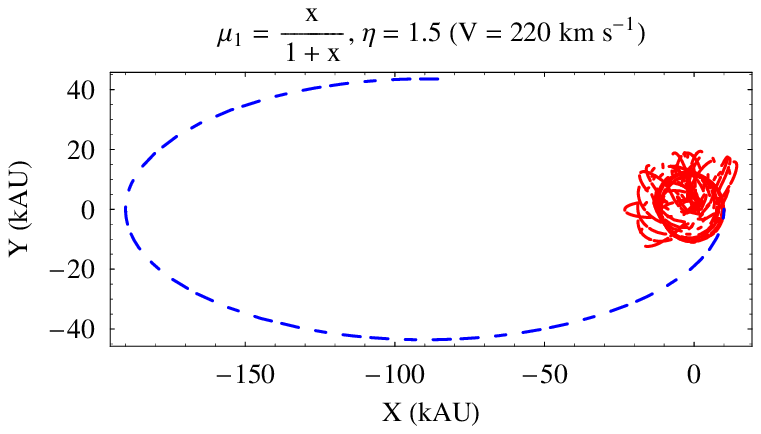}\\ %[height=0.3\textheight,width=0.3\columnwidth]{xz_polar_mu2_old.eps} &
\end{tabular} \par }
\caption{Numerically integrated orbits of an ecliptic Oort comet with $a=100$ kAU, $e=0.9$, $P_{\rm b}=31.6$ Myr. Dashed blue line: Newton. Dash-dotted red line: MOND with $\mu_{1}$. Upper panel: $\eta=2.0$ corresponding to $V=254$ km s$^{-1}$. Lower panel: $\eta=1.5$ corresponding to $V=220$ km s$^{-1}$. The initial conditions are $x_0=a(1-e), y_0=z_0=0,\dot x_0=0,\dot y_0=n a \sqrt{\rp{1+e}{1-e}},\dot z_0=0$. The time span of the integration is $-0.9 P_{\rm b}\leq t\leq 0$.\label{undici}}
\end{figure}

%\begin{figure}
% \includegraphics[height=.25\textheight]{icsipsilon_pazzo2}
%
%\caption{\footnotesize{Numerically integrated orbits of an Oort comet with $a=100$ kAU, $e=0.5$, $P_{\rm b}=31.6$ Myr. Dashed blue line: Newton. Dash-dotted %red line: MOND with $\mu_{1}$, $\eta=2.0$ ($V=254$ km s$^{-1}$). The initial conditions are $x_0=a(1-e), y_0=z_0=0,\dot x_0=0,\dot y_0=n a %\sqrt{\rp{1+e}{1-e}},\dot z_0=0$. The time span of the integration is $-3 P_{\rm b}\leq t\leq 0$.}\label{nove}}
%\end{figure}
%
%
The case $\eta=1.5$, yielding
\begin{eqnarray}
% \nonumber to remove numbering (before each equation)
  \mu_g &=& 0.60 \\
  L_g &=& 0.39,
\end{eqnarray}
is shown in the right panel of Figure \ref{nove} for $-P_{\rm b}\leq t\leq 0$.
%
%
%\begin{figure}
% \includegraphics[height=.25\textheight]{icsipsilon_pazzo}
%
%\caption{\footnotesize{Numerically integrated orbits of an Oort comet with $a=100$ kAU, $e=0.5$, $P_{\rm b}=31.6$ Myr. Continuous blue line: Newton. %Dash-dotted red line: MOND with $\mu_{1}$, $\eta=1.5$ ($V=220$ km s$^{-1}$). The initial conditions are $x_0=a(1-e), y_0=z_0=0,\dot x_0=0,\dot y_0=n a %\sqrt{\rp{1+e}{1-e}},\dot z_0=0$. The time span of the integration is $-P_{\rm b}\leq t\leq 0$.}\label{dieci}}
%\end{figure}
%
The case of highly elliptic orbits ($e=0.9$) is more intricate, as shown by Figure \ref{undici}.
%
%
%\begin{figure}
% \includegraphics[height=.25\textheight]{icsipsilon_pazzo_elli}
%
%\caption{\footnotesize{Numerically integrated orbits of an Oort comet with $a=100$ kAU, $e=0.9$, $P_{\rm b}=31.6$ Myr. Dashed blue line: Newton. Dash-dotted %red line: MOND with $\mu_{1}$, $\eta=2.0$ ($V=254$ km s$^{-1}$). The initial conditions are $x_0=a(1-e), y_0=z_0=0,\dot x_0=0,\dot y_0=n a %\sqrt{\rp{1+e}{1-e}},\dot z_0=0$. The time span of the integration is $-0.9 P_{\rm b}\leq t\leq 0$.}\label{undici}}
%\end{figure}
%
%
%
%
%
%\begin{figure}
% \includegraphics[height=.25\textheight]{icsipsilon_pazzo_elli2}
%
%\caption{\footnotesize{Numerically integrated orbits of an Oort comet with $a=100$ kAU, $e=0.9$, $P_{\rm b}=31.6$ Myr. Dashed blue line: Newton. Dash-dotted %red line: MOND with $\mu_{1}$, $\eta=1.5$ ($V=220$ km s$^{-1}$). The initial conditions are $x_0=a(1-e), y_0=z_0=0,\dot x_0=0,\dot y_0=n a %\sqrt{\rp{1+e}{1-e}},\dot z_0=0$. The time span of the integration is $-0.9 P_{\rm b}\leq t\leq 0$.}\label{dodici}}
%\end{figure}
%
%
%
%
%
%
The MOND paths resemble confuse clouds confined within a small spatial region.

The general features common to all the pictures shown may have consequences on the interaction of the Oort-like objects lying close to the ecliptic with passing stars \cite{Oo} by  reducing their perturbing effects and, thus, also altering the number of long-period comets launched into  the inner regions of the solar system, the number of comets left in the cloud throughout its history. Indeed, in the standard picture, the comets moving along very (Newtonian) elongated orbits may come relatively close to a star of mass $M_{\star}$ suffering  a  change in velocity $\Delta v$ which
approximately is \cite{Oo}
\eqi \Delta v = \rp{2GM_{\star}}{v_{\star}d},\eqf
where $v_{\star}$ is the star's velocity with respect to the Sun and $d$ is the distance of closest approach with the Oort object. Moreover, less elongated orbits would also reduce the perturbing effects of the Galactic tides.
\subsection{Nearly polar orbits}
\begin{figure}[!h]
{ \centering \begin{tabular}{c}
\includegraphics[height=0.3\textheight]{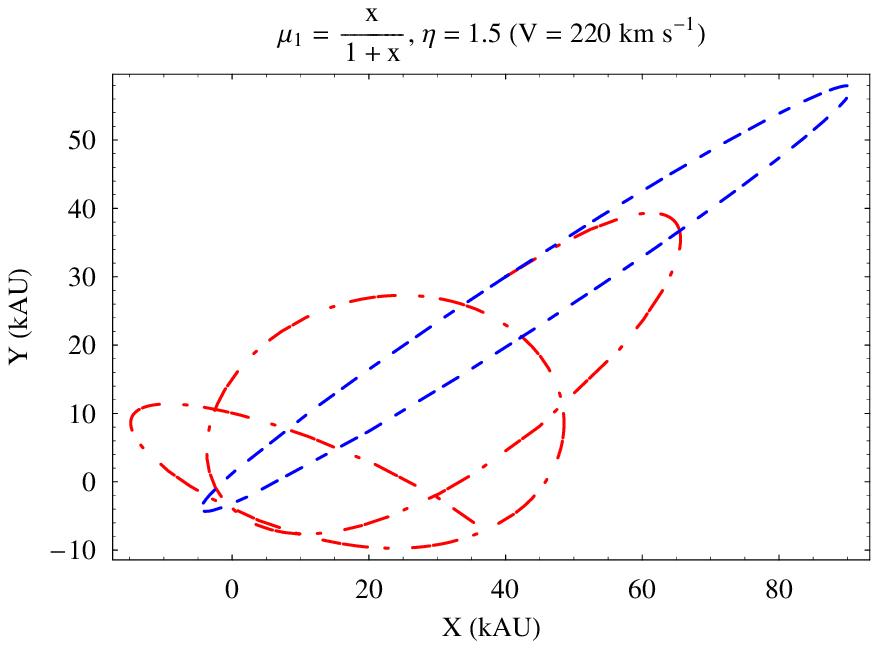}\\ %[height=0.3\textheight,width=0.4\columnwidth]{xy_polar_mu2_old.eps}&
\includegraphics[height=0.3\textheight]{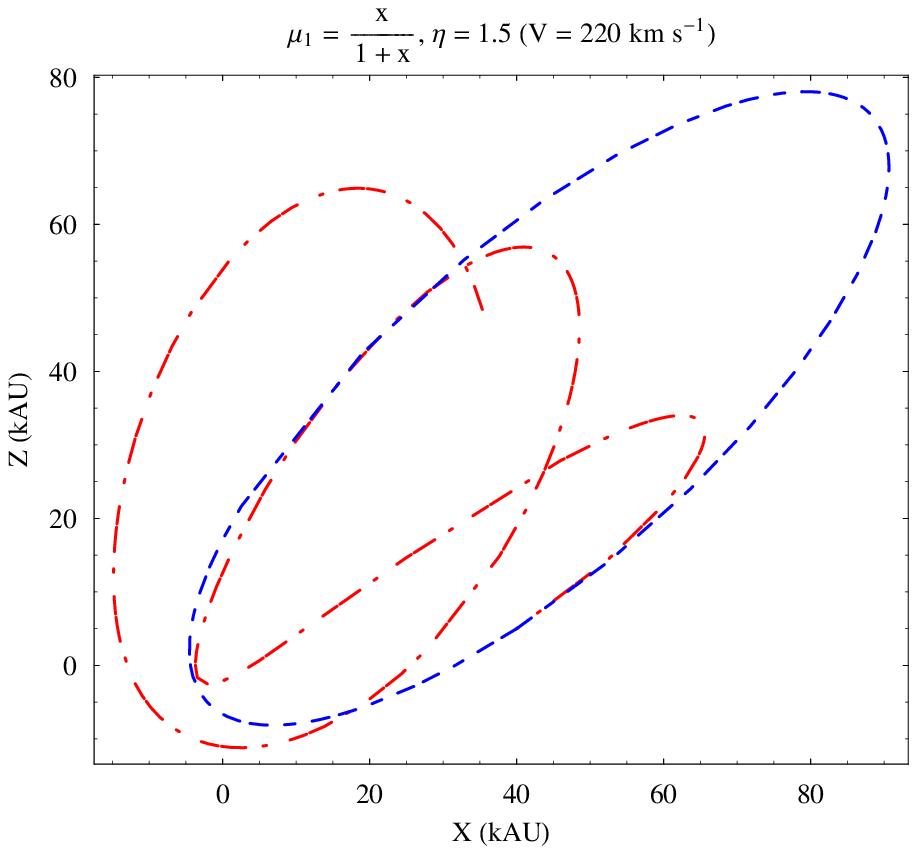}\\
\includegraphics[height=0.3\textheight]{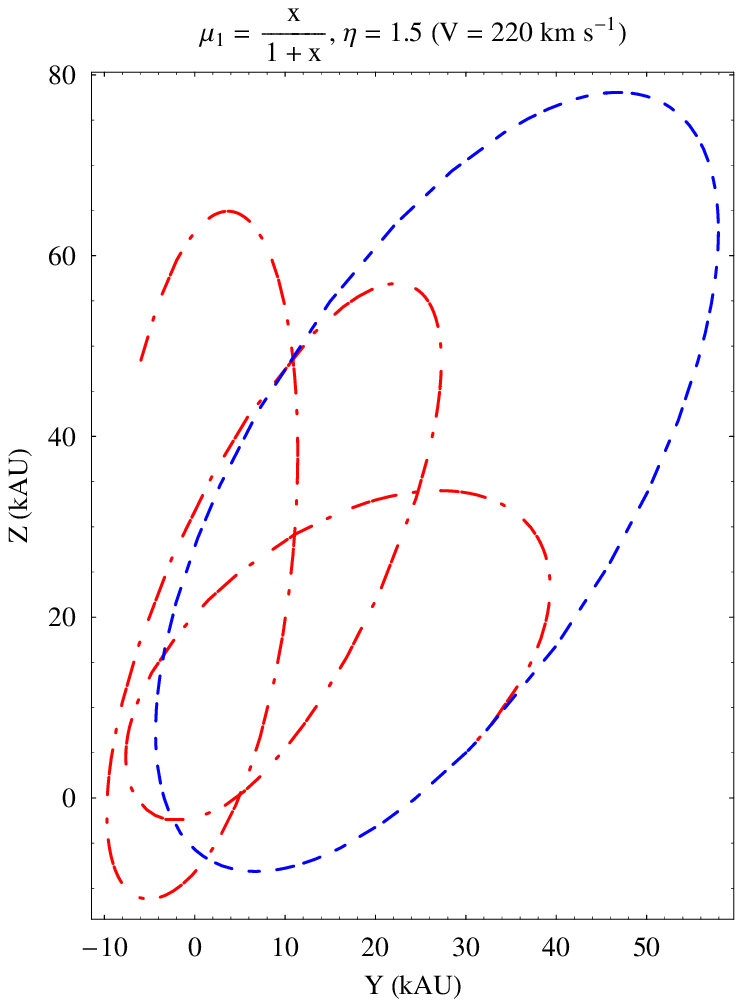}\\  %[height=0.3\textheight,width=0.3\columnwidth]{xz_polar_mu2_old.eps} &
\end{tabular} \par }
\caption{Sections in the coordinate  planes of the numerically integrated orbits of an Oort comet with $a=66.6$ kAU, $e=0.92$, $I=81$ deg. Dashed blue line: Newton. Dash-dotted red line: MOND with $\mu_{1}$, $\eta=1.5$ ($V=220$ km s$^{-1}$). The initial conditions are $x_0=40\ {\rm kAU}, y_0=30\ {\rm kAU}, z_0=5\ {\rm kAU},\dot x_0=-23\ {\rm kAU\ Myr^{-1}},\dot y_0=-15\ {\rm kAU\ Myr^{-1}},\dot z_0=-15\ {\rm kAU\ Myr^{-1}}$. The time span of the integration is $-P_{\rm b}\leq t\leq 0$.\label{kazza1}}
\end{figure}

Let us, now consider the case of orbits showing high inclinations $I$ to the ecliptic.
For space reasons we will only show some cases. In Figure \ref{kazza1} we depict the sections in the coordinate planes of an orbit with $a=66.6$ kAU, $e=0.92$, $I=81$ deg for $\mu_1$ and $\eta=1.5$.

\begin{figure}[!h]
{ \centering \begin{tabular}{c}
\includegraphics[height=0.3\textheight]{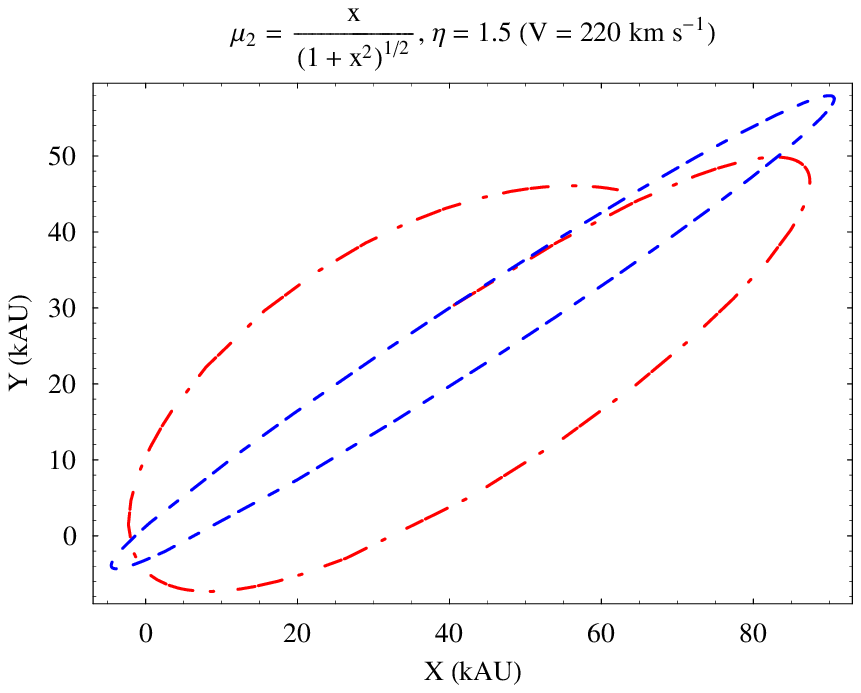}\\ %[height=0.3\textheight,width=0.4\columnwidth]{xy_polar_mu2_old.eps}&
\includegraphics[height=0.3\textheight]{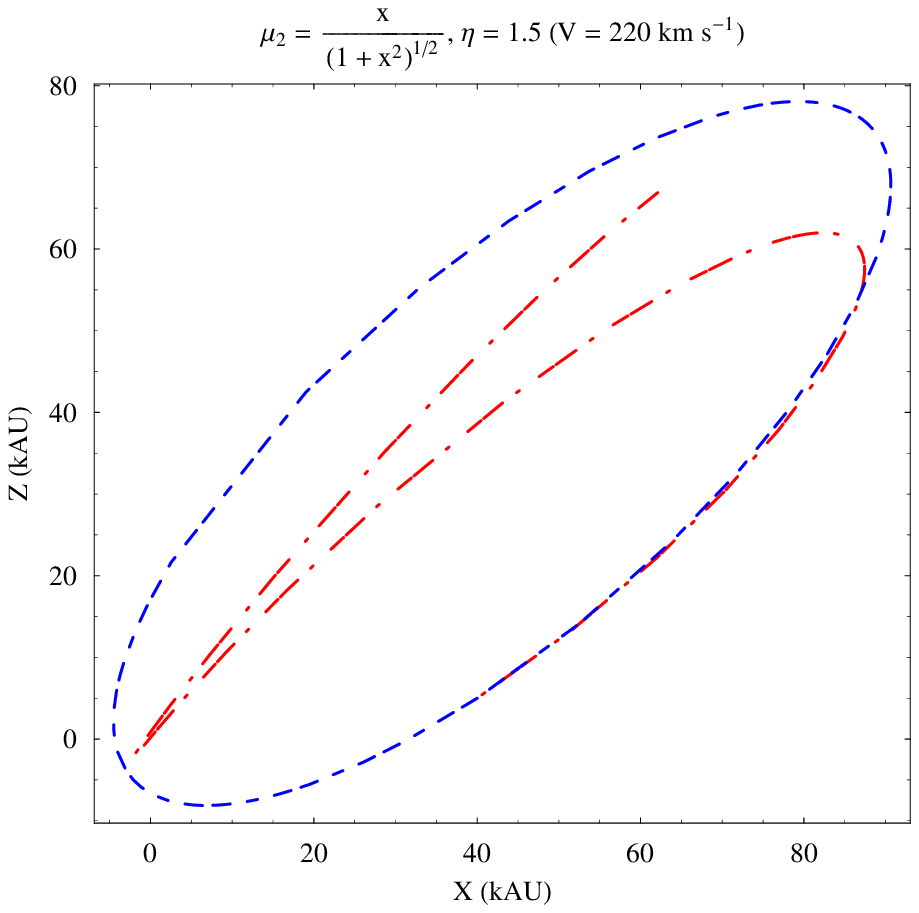}\\
\includegraphics[height=0.3\textheight]{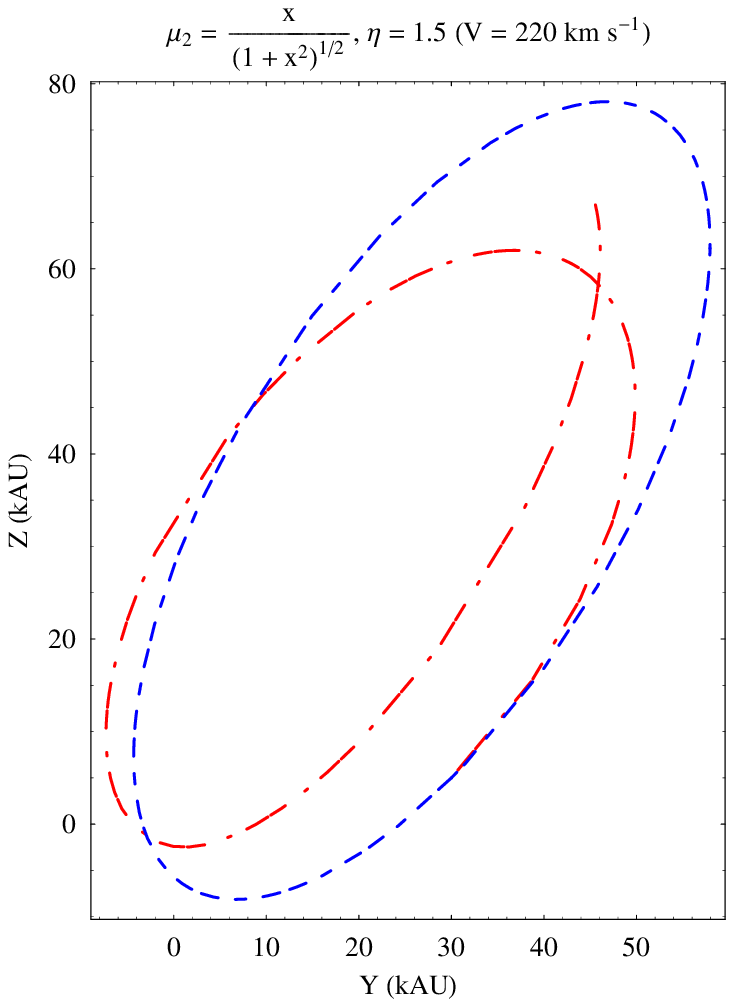}\\  %[height=0.3\textheight,width=0.3\columnwidth]{xz_polar_mu2_old.eps} &
\end{tabular} \par }
\caption{Sections in the coordinate  planes of the numerically integrated orbits of an Oort comet with $a=66.6$ kAU, $e=0.92$, $I=81$ deg. Dashed blue line: Newton. Dash-dotted red line: MOND with $\mu_{2}$, $\eta=1.5$ ($V=220$ km s$^{-1}$). The initial conditions are $x_0=40\ {\rm kAU}, y_0=30\ {\rm kAU}, z_0=5\ {\rm kAU},\dot x_0=-23\ {\rm kAU\ Myr^{-1}},\dot y_0=-15\ {\rm kAU\ Myr^{-1}},\dot z_0=-15\ {\rm kAU\ Myr^{-1}}$. The time span of the integration is $-P_{\rm b}\leq t\leq 0$.\label{pippa1}}
\end{figure}

The case of $\mu_2$ and $\eta=1.5$ is illustrated in Figure \ref{pippa1}.

It turns out that also in this case the MOND orbits are not closed, but the shrinking  is now less marked than in the ecliptic plane.

\section{Conclusions}
The structure and the dynamical history of the Oort cloud, subject to EFE in deep MONDian regime, may be  altered with respect to the standard Newtonian picture because highly eccentric orbits are not  allowed in the ecliptic plane by MOND  which, on the contrary, tends to strongly shrink them. As a consequence, one may speculate that the number of long-period comets launched in the inner parts of the solar system should be reduced because of the  less effective  perturbing actions of nearby passing stars, interstellar clouds and Galactic tides. Out of the ecliptic the situation is different because, although distorted with respect to the Newtonian case, the MOND orbits tend to occupy  larger spatial regions than in the ecliptic.

%%%%%%%%%%%%%%%%%%%%%%%%%%%%%%%%%%%%%%%%%%%%%%%%
%% BACKMATTER
%%%%%%%%%%%%%%%%%%%%%%%%%%%%%%%%%%%%%%%%%%%%%%%%

\begin{theacknowledgments}
I acknowledge the financial support received from INFN-Sezione di Pisa that allowed me to attend the Invisible Universe International Conference, 29 June- 3 July 2009, Paris.
\end{theacknowledgments}

%%%%%%%%%%%%%%%%%%%%%%%%%%%%%%%%%%%%%%%%%%%%%%%%
%% The bibliography can be prepared using the BibTeX program or
%% manually.
%%
%% The code below assumes that BibTeX is used.  If the bibliography is
%% produced without BibTeX comment out the following lines and see the
%% aipguide.pdf for further information.
%%
%% For your convenience a manually coded example is appended
%% after the \end{document}
%%%%%%%%%%%%%%%%%%%%%%%%%%%%%%%%%%%%%%%%%%%%%%%%

%%%%%%%%%%%%%%%%%%%%%%%%%%%%%%%%%%%%%%%%%%%%%%%%
%% You may have to change the BibTeX style below, depending on your
%% setup or preferences.
%%
%%
%% For The AIP proceedings layouts use either
%%%%%%%%%%%%%%%%%%%%%%%%%%%%%%%%%%%%%%%%%%%%

\bibliographystyle{aipproc}   % if natbib is available
%\bibliographystyle{aipprocl} % if natbib is missing

%%%%%%%%%%%%%%%%%%%%%%%%%%%%%%%%%%%%%%%%%%%
%% You probably want to use your own bibtex database here
%%%%%%%%%%%%%%%%%%%%%%%%%%%%%%%%%%%%%%%%%%%

%%%%%%%%%%%%%%%%%%%%%%%%%%%%%%%%%%%%%%%%%%%
%% Just a reminder that you may have to run bibtex
%% All of it up to \end{document} can be removed
%% if you don't like the warning.
%%%%%%%%%%%%%%%%%%%%%%%%%%%%%%%%%%%%%%%%%%%

\end{document}